# A Review on Blockchain Technologies for an Advanced and Cyber-Resilient Automotive Industry

**PAULA FRAGA-LAMAS[1], (Member, IEEE),**
**AND TIAGO M. FERNÁNDEZ-CARAMÉS[1], (Senior Member, IEEE)**
[1] Department of Computer Engineering, Faculty of Computer Science, Campus de Elviña s/n, Universidade da Coruña, 15071, A Coruña, Spain.
(e-mail: tiago.fernandez@udc.es; paula.fraga@udc.es)

Corresponding authors: Paula Fraga-Lamas and Tiago M. Fernández-Caramés (e-mail: paula.fraga@udc.es; tiago.fernandez@udc.es).

This work was supported in part by the Xunta de Galicia under Grant ED431C 2016-045, and Grant ED431G/01, in part by the Agencia Estatal de Investigación of Spain under Grant TEC2016-75067-C4-1-R, and in part by ERDF funds of EU under Grant AEI/FEDER, UE. Paula Fraga-Lamas would also like to thank the support of BBVA and the BritishSpanish Society Grant.

**ABSTRACT** In the last century the automotive industry has arguably transformed society, being one of the most complex, sophisticated and technologically advanced industries, with innovations ranging from hybrid, electric and self-driving smart cars to the development of IoT-connected cars. Due to its complexity, it requires the involvement of many Industry 4.0 technologies, like robotics, advanced manufacturing systems, cyber-physical systems or augmented reality. One of the latest technologies that can benefit the automotive industry is blockchain, which can enhance its data security, privacy, anonymity, traceability, accountability, integrity, robustness, transparency, trustworthiness and authentication, as well as provide long-term sustainability and a higher operational efficiency to the whole industry. This review analyzes the great potential of applying blockchain technologies to the automotive industry emphasizing its cybersecurity features. Thus, the applicability of blockchain is evaluated after examining the state-of-the-art and devising the main stakeholders' current challenges. Furthermore, the article describes the most relevant use cases, since the broad adoption of blockchain unlocks a wide area of short- and medium-term promising automotive applications that can create new business models and even disrupt the car-sharing economy as we know it. Finally, after a Strengths, Weaknesses, Opportunities, and Threats (SWOT) analysis, some recommendations are enumerated with the aim of guiding researchers and companies in future cyber-resilient automotive industry developments.

## I. INTRODUCTION

The automotive industry is one of the most technologically advanced industries with innovations ranging from hybrid, electric and self-driving smart cars to the Industrial Internet of Things (IIoT) integration in the form of IoT-connected cars. Under the Industry 4.0. paradigm [1], which represents the next stage in the digitalization of the sector, the automotive industry is facing operational inefficiencies and security issues that lead to cyber-attacks, unnecessary casualties, incidents, losses, costs and inflated prices for parts and services. Such issues are currently passed on to the different and heterogeneous stakeholders (i.e., individual and corporate car owners, service users, logistic businesses' clients or end customers) involved in the vehicle lifecycle. Industry 4.0 harnesses the advances from multiple fields, which allow for the massive deployment of sensors, the application of big data techniques, the improvements in connectivity and computational power, the emergence of new machine learning approaches, the development of new computing paradigms (e.g., cloud, fog, mist and edge computing), novel human-machine interfaces [2]–[4], IIoT enhancements [5] or the use of robotics and 3-D/4-D printing. The increasing capabilities offered by complex heterogeneous connected and autonomous networked systems enable a wide range of features and services, but they come with the threat of malicious attacks or additional risks that make cybersecurity even more challenging. In scenarios where the controlled systems are vehicles or vehicle-related systems, public safety is at stake, therefore strong cybersecurity becomes an essential requirement.

According to a Frost & Sullivan forecast regarding near-future investments [6], automotive IIoT spend is bound to increase from $ 12.3 bn in 2015 to $ 36.7 bn in 2025 at a Compound Annual Growth Rate (CAGR) of 11.5 %. In addition, the digital retailing in automotive IoT spending will increase at a CAGR of 29.1% from 2015 to 2025 and data driven business models will grow to a CAGR of 35% from $ 524.4 mn in 2015 to $10.5 bn in 2025. Automotive ICT spending is expected to increase from $ 37.9 bn in 2015 to $ 168.7 bn in 2025 with a CAGR of 16.1% due to new digitization initiatives that will include pilot software projects that will involve automotive Original Equipment Manufacturers (OEMs) and Tier 1s. Furthermore, OEMs digital transformation strategy roadmap is to currently develop digital services and move to a Car as a Service (CaaS) business model in the 2020s to then develop a Mobility as a Service (MaaS) model to eventually position the vehicle as an element of the future connected living solutions by 2030s.

In this context, blockchain technologies represent nowadays a move in the evolution of the Internet, enabling the migration from the 'Internet of Information' to the 'Internet of Value' and the creation of a true peer-to-peer sharing economy [7], [8]. According to a World Economic Forum survey report, 10 % of the worldwide Gross Domestic Product (GDP) will be stored on a blockchain by 2027 [9]. Considering also the prospects of the automotive ecosystem, blockchain technology can offer a seamless decentralized platform where information about insurance, proof of ownership, patents, repairs, maintenance and tangible/intangible assets can be securely recorded, tracked and managed. The ensured integrity of ledgers is one of the main aspects when dealing with transactions between the participants of the automotive industry. Their accuracy and immutability is essential for enforcing real-life contractual relations, avoiding poor practices and efficiently managing the supply chain. Furthermore, the ability to access verified data in real-time opens up a realm of opportunities and business models such as the automation of processes through Internet of Things (IoT) [10]–[14] and smart contracts, advances in predictive maintenance and forensics, smart charging services for electric vehicles, peer-to-peer lending, leasing and financing, or the introduction of novel models of collaborative mobility or MaaS.

Although a detailed description on the inner workings of blockchain technology is out of the scope of this paper, the interested reader can find detailed information in recent general reviews [15]–[23]. Specifically, a comprehensive overview on blockchain that emphasizes its application to IoT is provided in [24]. There is not much research work focused on the use of blockchain to enable cybersecurity.

For instance, in [25] the authors reviewed the main security issues that blockchain can tackle. Other works focused on specific security aspects. An example is presented in [26], where a cloud-based access control model is proposed. Other authors [27] focused on user identity management for cloud-based blockchain applications. Regarding the utilization of blockchain for specific applications, in [28] it is used to guarantee security and scalability in smart grid communications. Similarly, a Cyber-Physical System (CPS) [29] that makes use of a payment system based on reputation is presented in [30]. An interesting work is presented in [31], where a framework for fighting cyber-attacks when multiple organizations participate in information sharing is proposed. In the article, some game-based cyber-attacks are formally analyzed and validated through simulations. Finally, in [32], the authors review the use of blockchain and Content-Centric Networking (CCN) to ensure the security requirements for trusted 5G vehicular networks.

In contrast to the references previously cited, this work presents a holistic approach to blockchain for the automotive industry that includes both the basics for designing blockchain-based cyber-resilient applications and a detailed analysis on how to deploy and optimize blockchain technologies for the automotive industry. In addition, this paper is aimed at providing a global vision on how blockchain can transform the automotive sector radically and thus tackle part of its current challenges. The specifics of the blockchain implementation and other technical details of each use case are out of the scope of this article.

The rest of this paper is organized as follows. Section II reviews the most relevant security aspects involved in a blockchain-based development. Section III overviews the main issues and inefficiencies of the automotive industry and details a methodology for determining whether blockchain can help to tackle such issues. Section IV identifies scenarios where the automotive industry could leverage blockchain capabilities to enhance security, to reduce costs and to increase operation efficiency. Section V analyzes optimization strategies for designing blockchain-based automotive applications and studies their main challenges. Finally, Section VI is devoted to conclusions.

## II. BLOCKCHAIN BASICS FOR CYBERSECURITY

A blockchain is a distributed ledger based on a chain of linked blocks that enables sharing information among peers and that provides a solution for the double-spending problem [33]–[35].

Blockchain can provide multiple security benefits, which are detailed in the next subsections and are summarized in Figure 1, including the ones required by a cyber-resilient application: decentralization, cryptographic security, transparency and immutability.

### A. TAMPER-PROOF DATA

Any industry with different stakeholders needs a unique consistent data structure to read, update and take decisions

**IEEE** Access

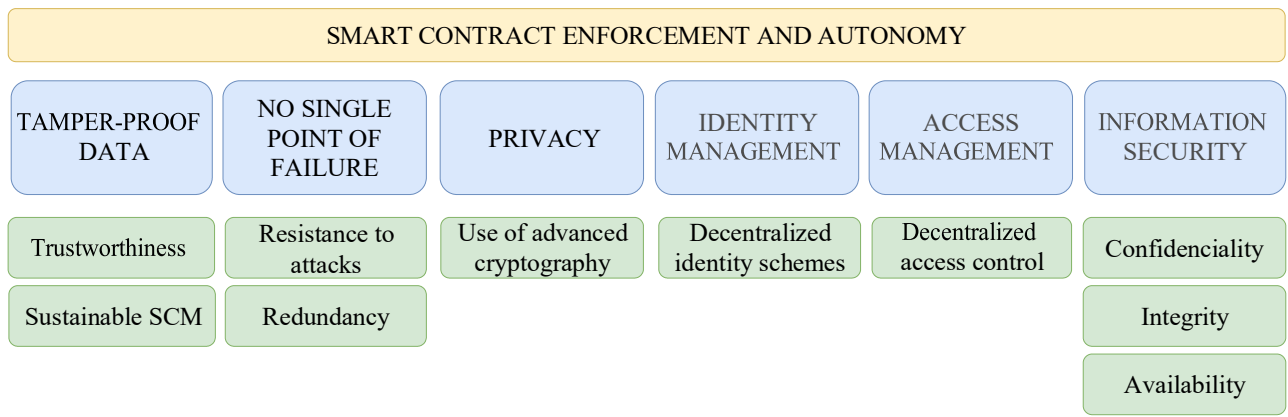

FIGURE 1: Blockchain key capabilities for cybersecurity.

[36]. Once a transaction is created in the blockchain, a new timestamp is recorded so that further modifications after such a timestamp will not be allowed. Traditional timestamping mechanisms rely on a trusted server that signs and timestamps the transactions with its own private key. Nevertheless, there is a risk: a malicious server might sign past transactions. Timestamping may be distributed, but then it can be prone to Sybil attacks [37], which blockchains like Bitcoin [38] prevent by linking blocks and using a Proof-of-Work (PoW) consensus mechanism. Another authors propose a decentralized timestamping service utilizing a similar concept of the long-term signature scheme standardized by ETSI [39] or a method to construct a secure and trusted timestamping authority [40].

### B. NO SINGLE POINT OF FAILURE

Blockchain performs data recording and storing using synchronous communication among the nodes through open-source sharing protocols. Open-source code has the advantage of being less prone to be altered by malicious parties, since it is monitored continuously by multiple contributors. However, like any other form of software, it can contain bugs and vulnerabilities.

Unlike traditional centralized databases, which store data in centralized clouds or server farms, a full blockchain node (a node of the blockchain that validates transactions) has a complete copy of the whole blockchain. This mechanism may derive into redundancy to some extent in specific scenarios, but the network becomes fault-tolerant and more reliable. In contrast, in cloud-centered architectures the cloud may become a single point of failure [41], since it can be unavailable due to multiple reasons (e.g., Denial of Service (DoS) attacks, maintenance tasks, software problems), and, therefore, the entire system may stop working. Moreover, only one single rogue node is required to alter the network performance through DoS attacks [42], eavesdropping or modifying the collected data [43]–[45]. To avoid the previously mentioned problems, a blockchain distributes its computing power among multiple nodes and, when a threat from a node is detected, the system is able to block its updates.

### C. PRIVACY

Blockchain uses public-key cryptography for providing security and privacy. Nowadays, there are two main public-key cipher suites for Transport Layer Security (TLS) [46], [47]: Rivest–Shamir–Adleman (RSA) based cipher suites [48] that also make use of RSA as the key exchange algorithm [49], [50]; and Elliptic Curve Diffie-Hellman Exchange (ECDHE), which is based on Elliptic Curve Cryptography (ECC) and performs exchanges through Ephemeral Diffie-Hellman [51]. Previous papers have already demonstrated that, in general, ECC is faster [52]–[55] and more energy efficient [56]–[62] than RSA. Nonetheless, in 2015 the National Security Agency (NSA) discouraged the use of Suite B, a set of cryptographic algorithms that made use of RSA and ECC. Apparently, the reason for such a statement was the fast evolution of quantum computing. In addition, National Institute of Standards and Technology (NIST) announced its plan to move forward to post-quantum schemes [63]. Recent developments in that way are described in [64]–[66] and in [67], where a cryptocurrency scheme based on Post-Quantum Blockchain (PQB) is defined.

It is also important to note that every user of a blockchain is identified by a public key or its hash. Although, to protect privacy, public keys are independent from the identity of a user, it is possible to determine certain identities by analyzing the performed transactions [68], [69], although such an analysis can be made more difficult by using multichains [70] or mixing protocols [71]–[73]. In addition, zero-knowledge proofs can be used for authentication, which enable proving that someone owns certain information without revealing it [74]–[77].

With respect to hash functions, they are essential for a blockchain, since they are needed for signing transactions. Therefore, hash functions should be fast and secure in terms of collision avoidance [78], [79]. Examples of such hash functions are SHA-256d, SHA-256 and Scrypt, which are already being used by multiple cryptocurrencies [80]–[85].

Finally, it is also worth noting that privacy has been recently considered as essential in different recent initiatives [86]–[88], which have suggested the use of techniques like ring signatures [89] or homomorphic encryption [90]–[93].

### D. IDENTITY MANAGEMENT

It is defined by the ISO/IEC [94] as the processes and policies involved in managing the life cycle and value, type and optional metadata of attributes in identities for a particular domain. Therefore, the identity provider controls the authorization of the different entities. Several approaches can be considered:

- ř Centralized schemes: the owner is a single entity that controls the system. It must be noted that their scope and utilization usually transcends this central organization (e.g., governments issue national identity cards valid for numerous entities).
- ř Federated schemes: the information, initially established in one security domain, can be utilized to access another domain (e.g., single sign-on schemes).
- ř User-centric schemes: the identity is owned and controlled by the single end-user (e.g., network anonymization).

For instance, decentralized identity schemes have emerged recently. Current strengths and challenges of applying DLT to identity management together with the evaluation of three proposals (i.e., uPort, ShoCard, and Sovrin) are analyzed in [95]. An example of implementation is illustrated in [96], where a permissioned blockchain with distributed identity management is used to increase security protection by rotating asymmetric keys.

An experimental cybersecurity cloud testbed with blockchain-based user identity management is described in [27]. The article includes experimental results of a penetration test in an Hyperledger application. Other authors [97] presented a cloud identity management solution to ease the creation of secure Infrastructure as a Service (IaaS) cloud federations. Other works focused on specific authentication schemes such as the proposed Horcrux protocol [98] that allows the end-users of self-sovereign identity to have the control of accessing their identities through a biometric authentication, or a cryptographic membership authentication scheme to support blockchain-based identity management [99].

### E. ACCESS MANAGEMENT

It represents the policies, processes and tools to identify, control and manage the authorized access to a system or application. For example, a system to control access and permissions through a blockchain is proposed in [100].

### F. INFORMATION SECURITY

Three main properties of the exchanged information should be preserved in order to consider it secure:

- ř Confidentiality. Unauthorized accesses should not be allowed to critical information. Therefore, the privacy of

data transactions should be protected. This is a problem in centralized storage systems, which are really common in finance or industry, since such an infrastructure can suffer attacks or internal leaks [101], [102]. To prevent such issues, blockchain decentralizes storage. Thus, if a node becomes compromised, the rest of the system should operate normally.

To preserve the confidentiality of a user, his/her private key has to be protected, because such a key is what is needed together with the user's public key in order to impersonate him/her. Key management systems like the one proposed in [103] can help to avoid key tampering.

Moreover, blockchain technology can also prevent IP spoofing and forgery attacks [41]. Furthermore, blockchain can help certificate authorities and support initiatives like Google's Certificate Transparency [104] in order to prevent fake certificates [105].

- ř Integrity. It prevents data modifications from unauthorized users. Moreover, it allows for recovering information modified by authorized users in case certain damage occurs.

Blockchains are conceived for storing data so that, once stored, it is very difficult to modify them. However, in very exceptional cases information can be altered by using hard forks, which originate a divergence from the previous version of the blockchain.

In the case of collecting information from third-parties (e.g., in financial or industrial processes), data integrity is essential, especially when such parties are not trusted beforehand. To solve this problem, some authors proposed a cloud-based framework for IoT devices that preserves information integrity with the help of a blockchain [106].

- ř Availability. It is the possibility of accessing the system data when needed. A blockchain guarantees the availability by distributing data among peers. However, in some scenarios, availability can be compromised through attacks. The most feared is the 51-percent attack (also called majority attack), where a single miner (i.e., a transaction validator) can control the whole blockchain and perform transactions at wish. In this case, although data are available, the availability for performing transactions can be blocked by the attacker. Obviously, data integrity is also affected by this attack.

### G. SMART CONTRACT ENFORCEMENT AND AUTONOMY

Effectively, a smart contract takes the terms of a traditional contract, encoding it up in the form of a business process and sharing it around the business network. Smart contracts are verified and signed when they are distributed across the business network. A smart contract is actually a piece of decentralized code that is stored on the blockchain and that runs autonomously when certain conditions are fulfilled. Therefore, there is no concept of reneging on a smart contract.

A smart contract can be regarded as an executable program that follows certain legal terms to manage physical or digital elements. Although smart contracts avoid issues related to human ambiguity, they do not depend on a state for their enforcement. Therefore, they are a mechanism to preserve performance on the deals of the parties involved.

In terms of legality, two different types of smart contracts can be distinguished: strong and weak. In contrast to weak smart contracts, strong smart contracts usually involve high revocation and modification costs. In addition, in the case of strong smart contracts, traditional law enforcers will be helpless after they are executed, since they cannot be stopped once initiated (either by involved parties or by a judge).

A smart contract can also be updated, so they need methods to add modifications that may be required legally. For instance, an online public database or Application Programming Interface (API) may be used to access the latest legal terms of the contract. Another method would consist in asking the involved parties to update the source code by themselves, what avoids depending on third-parties to perform such a task. To prevent one of the parties to modify a contract unilaterally, its terms may be defined as unmodifiable.

Although smart contracts are stored on the blockchain, they received data from external services called oracles that collect information from different sources. For instance, an oracle can monitor the status of an item in order to determine if it has arrived and write such a status on the blockchain. Then, the change on the status of the item could be detected by the smart contract, which can trigger the payment related to the purchase of the item.

There are different types of oracles depending on the collected data and on how they interact with their sources:

ř Software oracles handle online information. Examples of such an information could be the temperature of a stored product or the prices of purchased parts. The data originate mainly from web sources, like company websites. The software oracle extracts the needed information and pushes it into the smart contract.

ř Hardware oracles are designed to obtain data directly from the physical world. For example, Radio Frequency Identification (RFID) sensors in the supply chain industry. The biggest challenge for these hardware oracles is to report readings without sacrificing data security.

ř Inbound oracles insert information from the external world into the blockchain (e.g., an automatic buy if some asset hits a certain price).

ř Outbound oracles enable smart contracts to transmit data to the external world (e.g., a smart lock in the physical world which receives a payment on its blockchain address and unlocks automatically).

ř Consensus based oracles imply the combination of different oracles to determine the outcome of an event. Prediction markets like Augur [107] and Gnosis [108] rely heavily on a rating system for oracles to confirm future outcomes and to avoid market manipulation.

In practice, oracles are responsible for the correct execution of a smart contract, since the insertion of incorrect information may derive into an action that may not be reverted easily (e.g., certain money transfers). Due to this problem, several companies presented oracles that verify the collected data [109]. Recently, some blockchain-based applications have become more complex and involve the use of the concepts of smart contract, oracles and Decentralized Autonomous Organization (DAO). A DAO is a distributed application implemented to make it possible for multiple parties, humans or machines, to interact with each other [110]. The interaction between the members is arbitrated by a blockchain application that is controlled exclusively by a set of immutable and incorruptible rules embedded in its source code. A DAO can provide services or resources to third-parties, or even hire people to perform specific tasks. Hence, individuals can transact with a DAO in order to access its service or get paid for their contributions. DAOs are fully autonomous, as they do not rely on any central server and, therefore, they cannot be shut down randomly by any single party (unless their code was specifically designed for it).

Ethereum provides a programming language for distributed applications, but it is far from sufficient for complex DAOs [111]. Further research will be needed to explore new approaches to building DAOs with the appropriate standardization and interoperability [112].

In addition, it is still necessary to develop legal regulations to enforce smart contracts and resolve disputes properly. Only a few researchers have studied the problem of binding real-world contracts with smart contracts [113], as well as the issues that happen when the outcome diverges from the one demanded by the law [114]. Furthermore, the main security vulnerabilities of Ethereum smart contracts have already been analyzed in the literature [115], but there are still numerous issues to be further studied.

## III. EVALUATION OF THE NEED OF BLOCKCHAIN IN THE AUTOMOTIVE INDUSTRY

This section provides a comprehensive identification and classification of the current stakeholders of the automotive industry. Next, we introduce specific challenges of each stakeholder that can be faced by the use of blockchain. These challenges are then grouped into key management areas of interest. Finally, we present a flow diagram that can be used as a general guidance for deciding when it is appropriate to make use of blockchain and deciding its specific type.

In the automotive industry, wealth is created through transactions and contracts in business networks that generate a flow of goods and services. The underlying markets could include open markets such as a car auction, or a private market such as a supply chain financing. In every case, assets are transferred across the business network between the different stakeholders. There are mainly two different types of assets: tangible assets (e.g., a car) and intangible assets. Intangible assets can be subdivided into financial assets (e.g., instruments such as bonds); intellectual (e.g., a piece of intellectual property like a patent) or digital assets.

As it can be seen in Table 1, blockchain use cases can be structured into several categories across its two fundamental functions in the automotive industry: record keeping (static registry, identity and smart contracts) and transactions (dynamic registry or payments infrastructure).

After reviewing the current state of the automotive industry, it was decided to target stakeholders who are impacted by or can influence the outcome of a blockchain deployment. This includes customers, shareholders, internal and external stakeholders. Figure 2 represents the main analyzed stakeholders.

Moreover, after examining carefully their current role, the automotive business network and the strategic agenda of a number of platforms, projects and research programs [116]–[126], a list was compiled on the specific challenges of each stakeholder related to trust, transaction costs or other areas where blockchain can be applicable to face inefficiencies. The most important detected challenges are summarized in Table 2.

Furthermore, after analyzing the collected data, it can be concluded that most of the stakeholders face similar challenges and a joint strategy will be needed to maximize the impact of blockchain applications. As it can be seen in Table 3, challenge concerns can be grouped into three specific management areas: data, operations and finance:

1) Data management. A shared set of reference data between all the stakeholders is needed. Today, all the different stakeholders of the business network keep their own copy of the reference data and update it according to some procedure, maybe by e-mail or paper, when information changes. There is a need for a distributed record system that has to be used and shared across the business network. In this way, all the participants in the business network can have their own copy of the distributed ledger. Examples of these data could be a job card, an employment record or the tracking codes of a spare part.

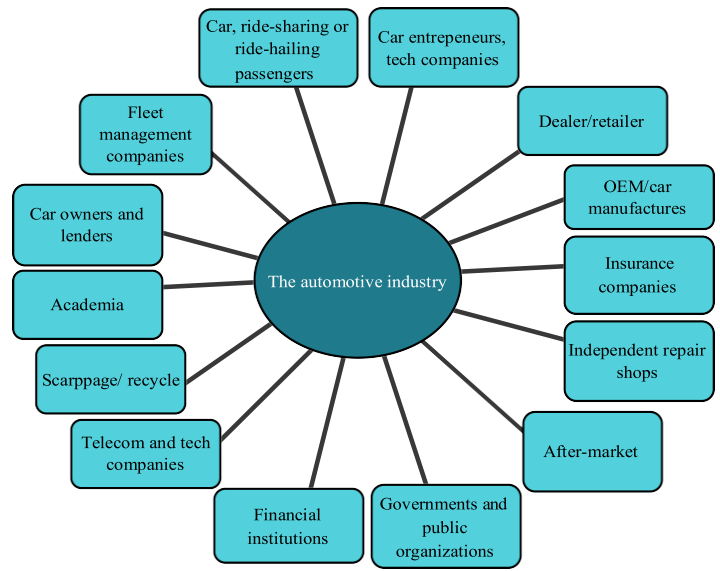

FIGURE 2: Main stakeholders in the automotive industry.

By putting all the information in a distributed ledger, it can actually be controlled who can change the data and who can actually get access to the data once they have been changed, thus making the whole process much more reliable.

Considering that the automotive industry spreads across different industries, countries and different regulatory boundaries, a shared set of data can be a very efficient way of managing reference data. The benefits imply reducing errors, improving real-time access to critical data and supporting natural workflows around creation, modification and deletion of the data elements.

Likewise, auditing (e.g., regulatory compliance) is a complex process, considering the fact that data and transactions are spread throughout many locations and are owned by many stakeholders. The fact that transactions are endorsed or validated by selected members of the business network has the effect of increasing the net trust within the business network. Furthermore, the fact that each member of the business network knows that they are sharing a common business process with the rest of the network also boosts trust.

When introducing a blockchain, privacy services control who can see what across the business network

| Category | Explanation | Use cases |
|---|---|---|
| Static registry | Distributed database for storing reference data | ⚊ Proof of ownership<br>⚊ Traceability<br>⚊ Patents |
| Identity | Distributed database with identity related information | ⚊ Identity fraud<br>⚊ Identity records |
| Smart contracts | Trigger automated and self-executing actions when predefined conditions are met | ⚊ Insurance-claim payout<br>⚊ Cash-equity trading |
| Dynamic registry | Distributed database that is updated with asset transactions | ⚊ Supply chain<br>⚊ Fractional investing |
| Payment infrastructure | Dynamic distributed database that is updated with payment transactions | ⚊ Cross-border payments<br>⚊ Peer-to-peer payments<br>⚊ Insurance claims |
| Several categories | Use cases composed by several of the previous groups Standalone cases not fitting in any of the previous categories | ⚊ Initial Coin Offering (ICO)<br>⚊ Blockchain as a Service (BaaS) |

TABLE 1: Main blockchain categories based on its main function.

| Stakeholder | Specific challenges |
|---|---|
| Car owners and lenders / buyers and sellers of pre-owned cars [127]–[129] | 1) Lack of transparency regarding the car's history<br>2) Unpredictable car maintenance and repair costs<br>3) Lack of trust in the outcome of maintenance and repair jobs<br>4) Absence of informed buying options<br>5) Absence of car insurance options<br>6) Lack of trust in autonomous vehicles and IoT-connected vehicles<br>7) High-level transactional experience to consumers whilst reducing the costs incurred by them |
| Fleet management companies / Car leasing or sharing (car-sharing, ride-sharing or ride-hailing) companies [120], [130] | 1) Lack of transparency regarding the car's history<br>2) Unpredictable car maintenance and repair costs<br>3) Lack of trust in the outcome of maintenance and repair jobs<br>4) Lack of interoperability with business partners<br>5) High operational costs, low margin<br>6) High costs in the car-sharing, ride-sharing and ride-hailing economy<br>7) Lack of trust in autonomous vehicles and IoT-connected vehicles |
| Car-sharing, ride-sharing or ride-hailing passengers [131], [132] | 1) More affordable car rides<br>2) Better maintained cars<br>3) Lack of trust in autonomous vehicles and IoT-connected vehicles<br>4) Lack of a common mobility provider platform<br>5) Lack of instant payment |
| Car entrepreneurs [122], [130] | 1) Expensive rates for car leasing and rental<br>2) Lower car-sharing, ride-sharing or ride-hailing partnership fees<br>3) Difficulties to set up business, unfair competition<br>4) Lack of trust in autonomous vehicles and IoT-connected vehicles<br>5) Lack of information sharing |
| Car dealers and retailers [122], [133], [134] | 1) Updated car ownership records<br>2) Updated repair and maintenance records<br>3) Updated purchase records<br>4) Lack of trust in autonomous vehicles and IoT-connected vehicles<br>5) Lack of information sharing |
| OEM / Car manufacturers and suppliers [118], [122], [134]–[137] | 1) Huge warranty claim costs<br>2) Enforcement of recommended maintenance and repair prices on the dealers<br>3) Customer complaints due to car dealers' violation of recommended maintenance prices set by car manufacturers<br>4) Lack of control of the car maintenance performed by authorized dealers<br>5) Weak customer loyalty<br>6) Cyber-attacks, system failure risks and enhanced security in autonomous vehicles and IoT-connected vehicles<br>7) Control of the logistics<br>8) Lack of information sharing |
| Insurance companies [138]–[140] | 1) Inflexible and non-customized policy pricing<br>2) 5-10% of all claims worldwide are fraudulent [143]<br>3) Costly and inefficient claim management<br>4) Inaccurate customer policy pricing<br>5) Lack of oversight over the quality and pricing for a collision repair |
| Independent repair shops [129], [134] | 1) Underutilized capacity<br>2) Customer retention<br>3) Low margins<br>4) Lack of brand confidence |
| After-market (producers, distributors and retailers of spare parts, garages) [134] | 1) Inefficient stock management<br>2) Market for counterfeit spare parts<br>3) Lack of transparency in warranty monitoring and enforcement<br>4) Low margins<br>5) Lack of brand confidence |

| Stakeholder | Specific challenges |
|---|---|
| Governments and public organizations [120], [124] | 1) Updated state registries (e.g., vehicle maintenance records, ownership rights, vehicle taxes, history of traffic fines)<br>2) Lack of trust in autonomous vehicles and IoT-connected vehicles<br>3) Compliance with the current legislation, particularly in terms of driver liability [117] or data protection<br>4) Enhanced interconnectivity with provision of open-source traffic and infrastructure data through a data cloud and willingness to shift to digital radio and universal network coverage.<br>5) Greater use of anonymisation and pseudonymisation in data collection and processing and provision of comprehensive information to vehicle owners and drivers about what data is collected and by whom.<br>6) Notifications of road conditions and traffic congestion in real-time<br>7) Trusted data for accident investigation and mitigating actions |
| Financial institutions [116] | 1) Updated car ownership records and insurance, maintenance and lien records on cars<br>2) Non availability of single reference point on all transactions |
| Telecommunication and tech companies, content and service providers [120], [141] | 1) Guarantee stable and secured Vehicle to Vehicle (V2V) and Vehicle to Infrastructure (V2I) communication to ensure efficient and safe vehicle coordination and cooperation<br>2) Lack of trusted connectivity among vehicles and between vehicles and infrastructure [117] |
| Scrappage/recycle and environmental groups [123] | 1) Control of greenhouse gas emissions<br>2) Full traceability of components<br>3) Long-term sustainability |
| Academia [121] | 1) Guarantee vehicle safety, security and autonomy<br>2) More efficient driving, development of optimized Human-Machine Interface (HMI)<br>3) Handle traffic management of highly and fully automated vehicles under mixed traffic conditions |

TABLE 2: Current specific challenges of the automotive industry that can be confronted using blockchain technologies.

(appropriate confidentiality between subsets of participants) and are also used to maintain this property of immutability across the blockchain, so the blockchain becomes tamper-proof. In a permissioned blockchain, it can be controlled who can see what parts (i.e., parts that are relevant to the stakeholders and their way of doing business) of the ledger. This creates a verifiable audit trail of everything owned/traded across the business network from the time it was created and put onto the blockchain. Such transactions cannot be altered, inserted or erased thanks to consensus, provenance and immutability, and the business logic actually embedded into the blockchain in the form of a smart contract.

2) Operations management is probably the most common cross-industry inefficiency considering the low implementation degree of the instruments of supply chain risk management [144]. SSCM [145] includes the complete traceability of the key assets, (i.e., a record when a car part is assembled or disassembled or is in the shipping process). Traditionally, traceability in the supply chain has been managed by using technologies like RFID [146]–[149], but blockchain technology, goes one step forward, enabling a new era of end-to-end transparency in the global supply chain system where stakeholders are able to share information rapidly and with confidence across a strong trusted network. Furthermore, the use of smart contracts provides a lower cost of transaction with a trusted contract monitored without the intervention of third parties. For instance, all the manufacturing data of where each part/asset of a subsystem has been in its journey from the manufacturer all the way through its integration into a car can be recorded. Note that this network can evolve with the shared set of referenced data to a more integrated and interlinked network of the different stakeholders.

Supply chain information can also include smart manufacturing processes (e.g., the individual computer-aided machine programming module that was used to create the part or other considerations), if they are relevant. Therefore, it ensures the traceability of an asset throughout its lifecycle. The advantages of this traceability are clear. Trust increases because it is possible to know who has owned each asset or where it has been, and hence, the whole supply chain becomes much more efficient.

It must be noted that if something goes wrong with a batch of cars or spare parts (i.e., a maintenance task or an insurance claim) diagnosing the incident or finding which subsystems or parts were actually involved can be easily solved, thus avoiding to perform a whole cross-fleet analysis or recall in the case of failure. As a result, including blockchain into a transaction processing system will derive in the following operational benefits:

- Transactions can be transformed to something that can take a number of days to almost real time.

| Management area | Stakeholders' challenges | Solutions | Benefits |
|---|---|---|---|
| Data | 1) Competitors and collaborators in a business network<br>2) Each business partner keeps their own database and forwards requests to a central authority for data collection and distribution<br>3) The owner of each information subset can be one or several organizations<br>4) Access to all the transactions over a specific reporting period is needed<br>5) Sensitive information must be continuously sent to the insurance companies<br>6) Some stakeholders (e.g., end-user) lack control over the exchanged data | 1) Shared set of referenced data<br>2) Each business partner maintains its own system (e.g., specific codification) within the blockchain network<br>3) There is a single view of the complete dataset in the business network<br>4) Privacy ensures only authorized user access | 1) Consolidated, immutable and consistent dataset with reduced probability of errors<br>2) Near real-time access to data<br>3) Interlinked network where code updates and transactions between stakeholders are naturally supported<br>4) Private sensitive data is shared on demand |
| Operations | 1) The provenance of each asset is hard to track<br>2) Traceability information: manufacturer, production date or batch data | 1) Sustainable Supply Chain Management (SSCM)<br>2) Complete provenance and traceability details of each critical asset<br>3) Data are accessible by each stakeholder (e.g., manufacturers, suppliers, car owners, insurance companies, government regulators) | 1) No single authority is the guarantor of provenance therefore, trust increases<br>2) More efficient ledger utilization<br>3) Individual and specific rather than cross-fleet information |
| Finance | 1) Financial data is dispersed throughout many systems with different characteristics and geographical locations<br>2) Letter of credits of a wider range of clients<br>3) Time and cost constraints | 1) Common ledger for letters of credit<br>2) Counterparties have the same validated record of transaction<br>3) Transaction records collected from diverse financial sources<br>4) A financial audit trail created with tamper-proof data | 1) Increased speed of execution<br>2) Reduced cost<br>3) Reduced uncertainty and risks (e.g., fluctuations in currency exchange rates)<br>4) Added-value applications (e.g., incremental claim payments)<br>5) An auditor is able to trace the accounting information from the blockchain to the source document (e.g., an invoice, a receipt, a form, a voucher) and view the complete process of a given transaction |

TABLE 3: Confronting today stakeholders' challenges.

⋎ Overheads and cost intermediaries that do not provide added-value can actually be taken out, making more efficient the whole business network. The distributed ledger and privacy services are used to manage the elements of the blockchain, therefore reducing the risk within the business network of tampering, fraud, or cyber-attacks. Furthermore, a net improvement of trust within the business network can be achieved because everyone uses the same way to keep their ledgers updated and their business processes flowing.

3) Financial management. In the automotive industry, these services involve letters of credit, financing, leasing, and cross-border import and export systems. Letters of credit are fundamental to the way that buying and selling occurs across borders. Numerous different individual documents are exchanged and signed by the banks and the different counterparties that represent the buyer and the seller in the business network.

Furthermore, automotive financing includes some verification steps (e.g., review of documents, scoring the risk or loan approval) that smart contracts can ease, therefore enabling to automatically negotiate payment on a car lease without a middleman.

Financial and logistics operations can be coupled with IoT devices. For example, when a pallet of goods actually crosses through an RFID reader into a warehouse. The seller could draw down a certain percentage of the letter of credit because through this RFID event, it can be ensured automatically that the goods actually made it part of the way to the end customer as well as the condition of the goods (e.g., if the assets were delivered in the agreed conditions of humidity, tilt or other parameters). These automated processes reduce the time of execution to almost real time. Therefore, they vastly reduce costs and risk for both the seller, the buyer and the correspondent banks who are involved in the process. This process could be applied to a number of other financing, and cross-border import and export systems.

Beyond the hype of Distributed Ledger Technology (DLT) technologies, it must be noted that in a trustful scenario or when stakeholders can trade directly, traditional databases or ledgers based on Directed Acyclic Graphs (DAG) [150] may be a better solution for daily operations. Certain industrial processes are inherently better suited for blockchain solu-tions. For example, financial services and governments core functions are clearly aligned with blockchain capabilities [151]. Specifically, Figure 3 shows a simplified flow diagram that can be used as a general guidance for deciding when it is appropriate to make use of blockchain technologies and to determine the type needed in a particular application. Further details on the specifics of the different types of blockchain can be found in [24].

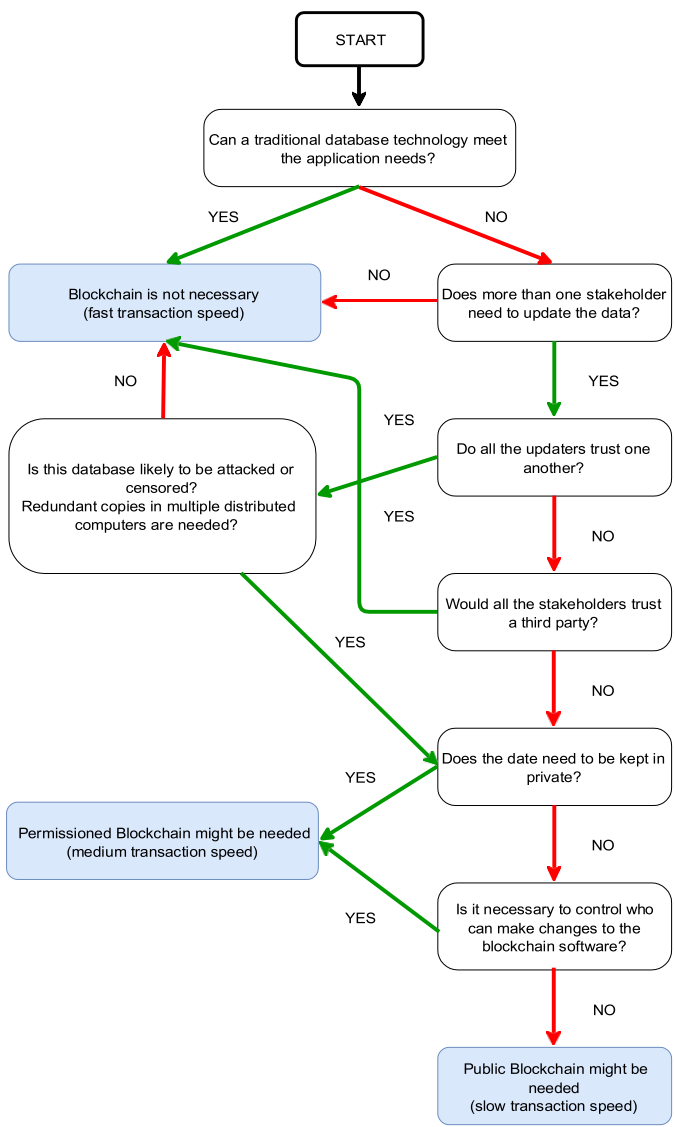

FIGURE 3: Flow diagram to determine the need of blockchain technologies in a specific application.

## IV. ADVANCED BLOCKCHAIN-BASED COMPELLING APPLICATIONS
This section reviews the most relevant blockchain applications for automotive environments (Figure 4).

- ☞ **Extended global vehicle ledger**
  A ledger that securely stores, updates, traces and shares data (e.g., car's maintenance, ownership history) in real time. Manufacturers can partner with a blockchain service provider to create a unique ledger among the network of OEMs to address logistics monitoring and control (e.g., issues related with spare parts quality and authenticity). The ledger can gather information about cars' history from different sources and even charge users to access the data [152]. The platform could be extended to receive payment for the rendered services (e.g., repairing a vehicle, or purchasing/selling vehicle data from/to a third party).

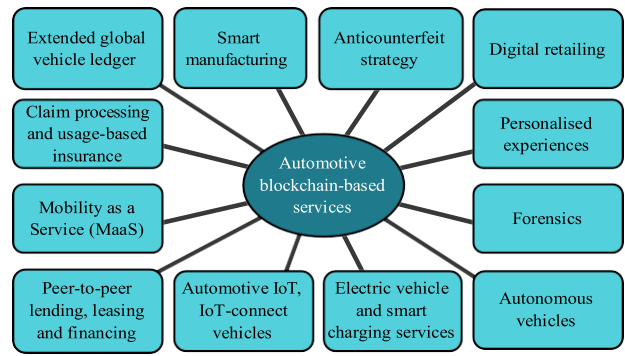

FIGURE 4: Automotive blockchain-based services.

- ☞ **Smart manufacturing**
  The inclusion of blockchain in software-based manufacturing can increase productivity and quality control, reducing the costs for tracking in inspections (e.g., it simplifies version management), warranty, inventory management [153], [154], ownership issues, maintenance or recycling tasks.

  A blockchain can also be used in a digital twin [120], [155], which represents digitally a physical asset in order to monitor its current state and to recreate its past and future [156]. In the automotive industry, assets (e.g., vehicles, tools, parts) may send data and notify events to its digital twin during their lifecycle. Thus, blockchain can be used to store securely all the mentioned information.

  An example of implementation was presented in July 2017 by Groupe Renault [157], which released a prototype created with Microsoft and VISEO to connect each car maintenance book to the vehicle's digital twin. These data are tamper-proof, fully traceable and visible to authorized parties such as the vehicle owner.

- ☞ **Anti-counterfeiting**
  Blockchain and IoT can provide an effective way to avoid fraud. On the one hand, counterparties can update the status of the items from the source to the point of sale, or even in some cases the whole lifecycle. On the other hand, sensors can be added to assets (e.g., to each part pallet shipped from the Original Equipment Supplier (OES)) to track their real-time location and status (e.g., that the shipment complies with the Estimated Time of Arrival (ETA)). It must be noted that this strat-egy will imply an extensive level of cooperation among automotive stakeholders and software developers.

Regarding odometer fraud, a solution that uses an in-car connector can be proposed to send vehicle mileage data to its digital logbook on a regular basis. If tampering is suspected, the displayed mileage can be compared with the recorded via an app. Furthermore, a car owner can log its mileage on a blockchain and receive a certificate of accuracy that could be used for guaranteeing selling conditions. For example, Bosch and TÜV Rheinland (a German certification authority) are collaborating to prevent the widespread practice of odometer fraud through a digital logbook solution [158].

ⅰ **Digital retailing and customer personalized experience**
Loyalty and reward programs can serve as customer incentives. In this use case a blockchain and smart contract-based solution can record customer purchases and issue loyalty points that can be used as a currency within the stakeholder loyalty network. The points are visualized and updated (e.g., redeemed as a discount) instantly for the whole network.

ⅰ **Claim processing and usage-based insurance**
Claims, particularly with complex insurance instruments, involve multiple parties. Nowadays, in the event of an accident, the liability is largely attributed to the driver, but autonomous vehicles need the consideration of other entities in the automotive ecosystem such as auto manufacturers, software providers, service technicians or vehicle owners. For instance, the insurance cost for a driver may be reduced by granting insurance companies access to driving data to demonstrate safe driving habits. In addition, certain collected information like braking patterns and speed may be used to avoid frauds.

The system would work as follows. First, the insurance company would create a public and a private key for every car, as well as a personal account stored in a cloud. The personal account is required by the company to know the actual identity of the policyholder. The public key would be stored into a secure database. The public and private keys would be used by the vehicle for every subsequent transaction with the insurance company. Thus, the vehicle stores in the cloud information on driving patterns that would be used by the insurance company to provide services. Certain critical information (e.g., vehicle location) could be stored in a blockchain in the in-vehicle storage. In case of an accident, the vehicle might fill a claim automatically by sending the information to the insurance company.

The vehicle owner may discontinue its contract with the insurance company or sell its vehicle. In such cases, the insurance company would remove the account from the cloud storage, so the vehicle would not receive further services.

ⅰ **MaaS**

Emerging technologies have created a new 'As-a-Service' business model in which initiatives such as Car Next Door [159] are growing fast. A blockchain-based platform would enable the interconnection of IoT-connected vehicles, autonomous vehicles, car-sharing, ride-sharing or ride-hailing providers and end-users to create a solution that records and executes agreements and monetary transactions to allow vehicle owners to monetize trips. Data (e.g., cost per mile, keys to unlock the car, insurance details, payment/billing details, information about vehicle owners, drivers and passengers) would be exchanged in a secure, reliable and seamless manner. The connections between the involved parties would be secured in order to protect their privacy (e.g., there would be no link between the actual user's identity and his/her route) and any unauthorized accesses to the vehicle (i.e., only authorized users would be able to locate, to unlock and to use a specific car). Furthermore, the platform could process all the payments upon completion of the trip and update the user's record with a history of the trip performed.

It is worth mentioning as an example an initiative of the Toyota Research Institute [160], which is exploring together with the MIT Media Lab the development of a new mobility blockchain-based ecosystem fostering the use of open-source software tools.

ⅰ **Peer-to-peer lending, leasing and financing**
Peer-to-peer models offer a business model that connects the involved entities and performs Know Your Customer (KYC) checks prior to leasing a vehicle, stores the leasing contract and automates the payment. Blockchain platforms will leverage secure communications and eliminate data risks. The extracted data can be used for analytics and for monitoring consumer behavior (KYC) in car leasing or rental. A couple of initiatives have studied the mentioned scenarios. For instance, in 2015 Visa and Docusign implemented a blockchain for a car leasing pilot service [161]. Similarly, Daimler AG and Landesbank Baden-Württemberg (LBBW) [162] made use of blockchain to perform financial transactions in a pilot project for monitoring capital market transactions and financial processes.

ⅰ **Connected services**
Vehicle owners can purchase infotainment or added-value services (e.g., parking, tolls) in a seamless manner based on pre-defined contracts that are stored and executed on the blockchain. For example, Carewallet [163] is a platform that allows for a full end-to-end integration of mobility services, vehicles and infrastructure.

Another application would be the introduction of blockchain for conventional wireless remote software updates. Nowadays, this is a centralized and non-scalable process with a partial participation of the supply chain (i.e., it does not include all the way from a service provider to a service center). Furthermore, there are potential privacy issues, since a direct link between the vehicle and the OEM can compromise the driver's privacy (e.g., its behavior or location) and only an OEM can verify communications or the history of update downloads. The use of a blockchain would imply an end-to-end distributed data exchange that involves service providers, OEMs, vehicles, service centers or assembly lines, and it will guarantee the user's privacy and the updated history, as well as the public verification of the authenticity of the software.

ɤ **Automotive IoT and IoT-connected vehicles**
Vehicles are becoming interconnected Cyber-Physical systems (CPSs) [164]. These CPSs have special-purpose sensors, control units (Electronic Control Unit (ECU) and On-Board Unit (OBU)) and wireless adapters to monitor their operations and communicate with their surroundings (e.g., Road Side Unit (RSU)) [165]). The penetration of the IoT paradigm in vehicles enables the collection of a huge amount of data. For instance, most vehicles manufactured in the last decade have On-Board diagnostics (OBD) ports, which are used for retrieving vehicle diagnostics. Another major development is the deployment of an Event Data Recorder (EDR) to store incident data based on triggering events (e.g., drastic speed reduction). Sensors and devices connected over a defined mobile network will enable the collection of data like driving events (e.g., mileage, speed), safety events (e.g., spare part replacement warning), maintenance events (e.g., annual service) and will be able to send these data to a ledger shared among the stakeholders (including the owner).

IoT applications help to monitor and control devices remotely and create new insights from real-time data. IoT, together with blockchain, can help to track, process and exchange transactions among connected devices. An example of intelligent communication between vehicles is proposed in [166]. Other authors [167] presented a lightweight scalable blockchain solution to face the challenges of traditional security and privacy methods in IoT-connected cars: centralization, lack of privacy or safety threats.

ɤ **Electric Vehicle and smart charging services**
Electric vehicle industry is growing in parallel with the demand for charging infrastructure. The connection of electric vehicles to the owner's smart home [168] and/or smart devices could lead to advanced services. For instance, the charging procedure might be customized according to the user personal habits (e.g., through the personal calendar). Such data could be used to guarantee that the vehicle is fully charged when needed. Furthermore, it also enables to choose the cheapest or more convenient charging cycle (e.g., avoiding peak load times).

A blockchain-based solution can be proposed for distributed accounting, for managing contracts or for automating billing and payments. Two scenarios could be considered: when the car owner charges the vehicle at a charging station owned by a third party or when the car owner discharges the electricity from the electric vehicle to the grid to support the stabilization of the energy network. The location and behavior of the user (e.g., using a specific charger on a specific day) could be tracked, but such a location information can remain private.

In the literature, there are some examples of implementations. For instance, a decentralized security model based on the lightning network and smart contracts is proposed in [169]. It involves registration, scheduling, authentication and charging phases. The proposed security model can be easily integrated with current scheduling mechanisms to enhance the security of trading between electric vehicles and charging piles. Another interesting example is described in [170], where a privacy-preserving selection of charging stations is presented.

ɤ **Autonomous or self-driving vehicles**
Since most of the car crashes are the result of human errors, a computer would be an ideal driver, as it can use complicated algorithms to determine appropriate driving measures. Autonomous vehicles are equipped with advanced IoT capabilities, navigation devices and computer vision technology to drive autonomously with limited or no human intervention. Leveraging blockchain as an underlying communication mechanism will guarantee trust and dependability on these systems. Furthermore, since cybersecurity is currently a main concern for autonomous and IoT-connected vehicles, the main threats and attacks to automated vehicles have been identified in [171]. For instance, another authors are focused on introducing peer-to-peer usage models. For example, Hasan et al. [172] propose a blockchain-based platform that can provide autonomous vehicles with share ride services.

ɤ **Forensics**
Forensics is becoming an important feature in a vehicle design and operational lifecycle. Interested stakeholders include insurance companies and law enforcement who are interested in crime solving (e.g., vehicle location information can be useful in a burglary or homicide) or crash incident investigations. In recent years, forensics has been further used by insurance providers and by companies giving vehicles to their employees for business-related activities.

IoT-connected and autonomous vehicles gather a huge amount of information that can be significant for manufacturers, service providers, drivers or insurance companies in case of an incident or accident. This capability to collect data within and around the vehicles can make a significant impact on the forensics field. The topics has to be further studied, but an example of permissioned blockchain forensic framework can be found in [139].

## V. BLOCKCHAIN IMPLEMENTATION AND DEPLOYMENT STRATEGY

### A. SWOT ANALYSIS

After analyzing blockchain technologies in Section II, this subsection evaluates its applicability based on a Strengths, Weaknesses, Opportunities, and Threats (SWOT) analysis that summarizes the main key issues that have to be considered when deploying blockchain technologies for the automotive industry.

### 1) STRENGTHS

As it can be observed in Table 4, blockchain brings numerous advantages. Its main strengths are operational efficiency and resiliency: by removing middlemen, transactions can be simplified and their cost can be lowered (e.g., banking fees).

Another strength is that smart contracts can be coded to perform autonomous transactions (e.g., decisions on business processes) based on data acquired by IoT devices or sent by different stakeholders. It must be noted that today, the data from the different stakeholders are stored in centralized databases, or even in paper, which implies costly and unreliable business processes. Moreover, these data are error prone and subject to hacking, unintentional errors or frauds as they go along the complex network of stakeholders. In contrast, the underlying technology behind blockchain (e.g., advanced cryptography) prevents the recorded data from being modified. Thus, records are irreversible and tamper-proof. Non-repudiation and immutability guarantee that there is a unique and historical version of the data that is agreed and shared among all the stakeholders (e.g., a shared set of referenced data).

Data transparency is guaranteed by providing global access to the blockchain. Since different stakeholders are able to upload information to the blockchain, it can become the storage of an enormous amount of trusted information that might be used for big data analytics.

Moreover, the fact that a blockchain can be replicated on every full node provides redundancy and guarantees that the stored data will resist unexpected events and cyberattacks.

For instance, full traceability, asset provenance and quality control on how parts or cars are stored, inspected and transported, can enhance accountability and give proof of ownership for all the involved parties. Therefore, relevant stakeholders can verify or inspect such an information at any time or at a specific moment, thus creating dynamic and fluid value exchanges.

### 2) WEAKNESSES

The major blockchain weaknesses are related to the immature status of the technology (e.g., lack of scalability, high energy consumption, low performance, interoperability risks or privacy issues). In the case of IoT-connected cars or infrastructure, smart contracts will be automatically executed and in some cases they will depend on the injection of source information from external oracles. Therefore, it is presumable that this scenario will be indeed appealing for criminal activity or malicious attacks.

In addition, nowadays, usability is another challenge (e.g., no intermediaries can be contacted in case of users' credentials loss) and the customer is usually not familiarized with the technology.

An additional weakness is the cryptocurrencies volatility, which can represent a limitation to the short-term adoption of blockchain-based payments.

Regarding the available development tools, they are still in an early stage and the adoption of common standards is still ongoing. Besides, it must be noted that developing blockchain-based applications require high-level specific technical skills and human resources are scarce and costly.

Finally, it is worth remarking that, in some cases, blockchain may not be the most suitable technology for a business use case or process (as previously discussed in Section III) and it is key to succeed the adaptation of corporate governance models to decentralized exchanges of value.

### 3) OPPORTUNITIES

In relation to opportunities for the automotive industry, blockchain allows for gaining industrial competitiveness, for entering into new markets or for developing new types of business models thanks to the use of DAOs and low transactions fees. Blockchain also represents an opportunity to reduce the information asymmetry that today exists among the different stakeholders.

Moreover, in the automotive ecosystem, blockchain can definitely help to prevent fraud and to reduce the possibility of a systemic risk (e.g., the risk of collapse of an entire market caused by intermediaries and/or idiosyncratic events).

Specifically, due to the network effect, when a high number of stakeholders are involved, blockchain-based supply chains can be more efficient, since data can be shared nearly instantaneously among different heterogeneous actors. Nonetheless, the impact of such big data-enabled applications depends on the amount and quality of the collected information.

The use of open-source code is also essential in order to increase security and transparency. It is important to note that, although this kind of code is still susceptible to bugs and exploits, it is less prone to malicious modifications from third parties, since it can be monitored constantly by any stakeholder.

For instance, in shipping processes the transportation and logistics sector rely on a global chain of actors including shipping lines, freight forwarders, port and terminal operators, and customs authorities. All of them constantly need to exchange information about the origin of goods, tariff codes, status, classification data, import/export certificates, manifests and loading lists. Nowadays, some paperwork necessary to process cross-border shipping is done manually and operational information is transmitted over the phone, e-mail or fax. Such processes are prone to errors, tampering and delayed communication. If inserted into a blockchain, the trading with external stakeholders can be eased by offering integrity, transparency, security and paperless flows of data that can greatly decrease the time and costs associated with current intermediaries, as well as reduce the verification processes in order to ensure the overall conformity and delivery.

| | Positive | Negative |
|---|---|---|
| | **Strengths** | **Weaknesses** |
| Internal | 1) Operational efficiency<br>2) Cyber resiliency<br>3) No need for intermediaries that do not provide added-value<br>4) Fast and simple transfers with low fees<br>5) Automated transactions by means of smart contracts, IoT enabler<br>6) Reduction in human errors<br>7) Accountability, verified, timestamped, and immutable auditable data<br>8) No data loss neither modified nor falsified data<br>9) Security and modern cryptography<br>10) Non-repudiation<br>11) Transparency<br>12) Global accessibility<br>13) Trusted big data analytics platform<br>14) Decentralization<br>15) Traceability, asset provenance<br>16) Dynamic and fluid value exchange<br>17) Accountability, proof of ownership and rights | 1) Immature, early stage of development<br>2) Scalability issues<br>3) High energy consumption<br>4) Low performance<br>5) Lack of interoperability<br>6) Privacy issues (in some scenarios)<br>7) Criminal activity, malicious attacks<br>8) Dependent on input information from external oracles<br>9) Poor user experience, customer unfamiliarity<br>10) In case of users' credentials loss (e.g., a wallet), no intermediary can be contacted<br>11) In specific use cases, subject to cryptocurrency volatility<br>12) Limitation of smart contract code programming model<br>13) Wallet and key management<br>14) High-skilled human resources (scarce and costly)<br>15) Complexity (blockchain concepts are difficult to be mastered)<br>16) Lack of trust in new technology suppliers<br>17) Core business use cases or processes may not be suitable for the use of blockchain<br>18) Poor corporate governance |
| | **Opportunities** | **Threats** |
| External | 1) Industrial competitiveness (e.g., reduced transaction costs, enhanced cybersecurity, full IoT automation)<br>2) Market diversification (e.g., supporting car sharing)<br>3) New business-model enabler<br>4) Rebalancing information symmetry between stakeholders<br>5) Fraud reduction<br>6) Reduced systemic risk<br>7) Network effect<br>8) A huge amount of heterogeneous data pushed into the blockchain by different actors for data analysis (big data applications)<br>9) Open-source code<br>10) Ease in cross-border trade<br>11) Reduction of verification procedures<br>12) Digital twin enabler<br>13) Circular economy enabler | 1) Perception of insecurity or unreliability<br>2) Technological vulnerabilities<br>3) Divergent blockchains, ledger competition<br>4) Low adoption from important stakeholders<br>5) Unfavorable government policies, legal jurisdiction barriers<br>6) Institutional adoption barriers<br>7) Medium or long-term investment<br>8) Not adequate for external customers, readiness for adoption |

TABLE 4: SWOT analysis for blockchain in the automotive industry.

Blockchain can also enhance the capabilities of a digital twin, which enables digital representations of physical assets to reflect reality through simulations based on information collected from IoT devices. Examples of such improved features can be traceability of electric and electronic devices along their lifecycle, the guarantee of the provenance and authenticity of components, the registration of events from initial product design and approval processes through manufacturing, the verification of the delivery process to customers and the corresponding after sale events, the inventory management using blockchain to validate signatures and orders, or even the submission of offers from different suppliers directly to a blockchain.

Finally, blockchain can also bring new opportunities to the circular economy by ensuring traceability, by providing incentives to recycle and by enabling trust-based reputation systems.

### 4) THREATS

With respect to threats, they are related to several factors. First, technology can be still distrusted by the market, since it can consider it as insecure or unreliable, mainly due to software problems or cryptocurrency volatility.

Code vulnerabilities in blockchain or smart contracts are a threat to a sustainable adoption and can damage brand reputation. An infamous example is the DAO attack of 2016, which exploited a combination of previously reported security vulnerabilities with a cost of around $ 50 million worth of Ether and a devaluation of the DAO by a third[115].

As it was previously mentioned, in some cases information can be altered by using hard forks. Although this kind of forks can happen for technical reasons to fix vulnerabilities, they can also result from regulatory interventions (e.g., different jurisdictions that take varied approaches to blockchain management) or even as a result of a divergence on the ledger itself in order to provide different features.

Another threat if the fact that some stakeholders may think that the proposed system is too complicated, so the adoption rate on a worldwide basis could be low.

It must be noted that unfavorable government policies, legal regulations and institutional adoption barriers slow down and threaten the mainstream adoption of blockchain. Potential barriers may arise to the use of smart contracts. A new subset of law, denominated as Lex Cryptographia [142], that includes rules governed through self-executing smart contracts and DAOs, will have to re-evaluate the interaction between four regulatory forces: the threat of law enforcement, the manipulation of markets (financial incentives and disincentives), social pressure and the centralized intermediaries (i.e., internet service providers). For instance, the jurisdiction of smart contracts is still under debate [114].

With respect to Return On Investment (ROI) aspects, it must be indicated that applications based on blockchain technologies are considered as medium or long-term investments and as not adequate for being integrated into every existing process. In fact, most current solutions are still in the prototype stage, but it is likely that more mature applications will reach a broad market in the next years.

Moreover, if blockchain technology becomes a practice, it can have an impact on a company relationship with their customers. However, some customers may refuse to adopt it, as they might still consider personal interaction important. In addition, despite investing in human capital in order to improve customer service, market share may be lost, since companies may start to compete in terms of pricing.

### B. FURTHER RECOMMENDATIONS

Despite the promising foreseen future of DLT, and specifically of blockchain technologies, the SWOT analysis of the previous Section revealed several challenges that may hinder their short-term development and deployment:

- Technical complexity: the scientific community is researching on scalability, privacy, security and post-quantum cryptography in order to face the main design limitations in transaction capacity, in validation protocols or in the design and implementation of smart contracts and DAOs. Moreover, it is necessary to introduce novel methods to foster decentralized approaches in business processes.
- Interoperability issues: the critical participants of the business network should be involved to guarantee the adoption of blockchain and its integration with third-party and legacy systems. To achieve full interoperability is necessary to adopt collaborative implementations and use international standards for trust and information protection (i.e., access control, authentication and authorization). For instance, Federated Identity Management (FIM) [173] is required to guarantee the authentication across multiple enterprises. At an international scale, such a FIM currently covers only a low Level of Assurance (LoA). The required LoA (from LoA 1 to LoA 4), as defined by the ISO/IEC 29115:2013 standard, is mainly based on the associated risks (probability of an event multiplied by its potential impact) derived from an authentication error and/or the misuse of credentials.

- Blockchain infrastructure: a comprehensive trust framework that can fulfill all the requirements for the use of blockchain must be created.
- Blockchain architecture: the design of the architecture should consider the company's decentralization requirements. In general, a private blockchain may be sufficient for the back-end. Private blockchains have been often undermined since the usage of a technology originally conceived to foster decentralization in a fully centralized way may be seen as a contradiction. Nevertheless, this type of blockchain is able to reduce the risk of data tampering and it can enable task automation.

In the scenarios where multiple organizations need to access data, such as most of the applications of the automotive industry, a consortium or federated blockchain may be preferable. They restrict user access to the network and the actions performed by the participants. This kind of blockchain can be maintained by nodes that belong to organizations of the consortium, and it could be used as a shared ledger. For instance, a public blockchain can be used when managing automatic payments with existing cryptocurrencies, or when there is a need for provide trust between organizations using an unmodifiable ledger.

- Standardization [174]–[179] and testing: after a deep understanding of the actors, supply chains, products, markets, services, and Key Performance Indicators (KPIs) involved in an automotive specific use case, all the operational and technical requirements have to be analyzed and agreed. As a first stage, when the blockchain is created, it should be tested in the field with the agreed criteria to verify if it works as needed. As a second stage, different indicators should be evaluated in terms of privacy, security, energy efficiency, throughput, latency, privacy, cost efficiency, blockchain capacity or usability, among others. For instance, considering the hyped state of blockchain, developers may fake their blockchain performance to attract investors (e.g., Initial Coin Offerings (ICO)), driven by the expected profits.
- Regulatory and legal aspects: the lack of a clear regulatory environment (e.g., decentralized ownership, contingencies in smart contracts, international jurisdiction, cross-border trade) and democratic-by-design models of governance are concerns that hinder the potential impact of blockchain. Companies in countries with supportive regulations will have a competitive advantage to develop innovative business models, that they will be willing to exploit legally. Furthermore, blockchain can enable value distribution models interoperable across organizations, improving the economic sustainability of both contributors and organizations.

ⱦ Organization, governance and culture: organizations' willingness and corporate governance will play an important role in the adoption of blockchain since coopetition (cooperative competition) and a collaborative mindset is required in order to engage all the stakeholders and adopt new ways for creating value.

For instance, in the collective imaginary, Bitcoin is frequently associated with fraud and pyramid or Ponzi schemes and it has been often misused to refer to blockchain. Therefore, there are still cultural barriers and misinformation that must be confronted.

ⱦ Suitable training and advisors: mastering the blockchain concepts requires a highly technical background that is necessary to fully realize the potential of the technology. Advisory boards should be comprised of influential leaders and experts in the areas of blockchain, crypto-technologies, IoT, cyber security, insurance, financial technologies, venture capital and business development. For instance, the usability of blockchain-related applications is still not adequate for the average user. Therefore, further efforts should be carried out to avoid the excessive underlying complexity.

ⱦ Business strategies (investments, acquisitions and partnerships): there are huge prospects of investment and new players entering the market, over 1,700 digital startups are aiming to disrupt the automotive industry. Technology companies and specialized startups will support OEMs and Tier 1s on their digital transformation in two main platforms: Business to Business (B2B) and Business to Customer (B2C).

Automotive companies will have to experiment with different blockchain projects in order to discover where the ROI/value resides or can be created (e.g., whether if there will be additional sources of revenues or profits, disruptive added-value services, cost savings, stronger brand image, cyber resilience, fraud reduction, improvements in customers' user experience). Nevertheless, in some scenarios the payoff may require that companies wait until blockchain solutions be more robust, scalable, interoperable or demand less custom development (i.e., long-term investment).

Another challenge is the development of standards considering that several blockchain-based systems may have to coexist within the automotive industry; likely, there will be many private permissioned blockchains, due to the business competitiveness, and multiple public blockchains. Therefore, organizations will be compelled to guarantee the interoperability between blockchains. To tackle these challenges, consortiums are now beginning to emerge; for example, on May 2018, the global consortium Mobility Open Blockchain Initiative (MOBI) [180] was announced to examine the potential of blockchain and distributed ledger technologies to create a novel digital mobility ecosystem more efficient, greener, affordable, safer, and more widely accessible. The consortium hopes to bring together automakers, suppliers, startups, technology firms, blockchain companies, NGOs, academia and government agencies.

ⱦ Network effect: the degree of industry adoption will determine the benefits of blockchain technology in the automotive industry, as the volume of exchanged information will increase. When the adoption reaches a critical mass, it could evolve into an industry practice. However, at the beginning it may be difficult to obtain stakeholder commitment considering the different levels of digital readiness and the difficulty of integrating legacy processes with novel systems and practices. For instance, an initial requirement is the recognition of the gains of a blockchain-based collaboration.

ⱦ No-one-size-fits-all solution: the adoption of blockchain unveils a broad area of short- and medium-term potential scenarios that could disrupt the automotive industry, as we know it today. Nevertheless, there is no one-size-fits-all technological solution for the automotive industry.

From the business standpoint, it can be assumed that small ventures (e.g., start-ups) will be more disruptive and take more risk than established companies. In the short-term, the greatest impact will come from the technology-driven transformation of global supply chains, although ultimately many other aspects will be affected. Blockchain's strategic value in the automotive industry would be focused mainly in operational efficiencies and cost reduction. The costs in existing processes can be optimized by removing unnecessary intermediaries or diminishing the administrative workload of record keeping and transaction reconciliation (e.g., speeding up claim processing). The blockchain-based processes can capture lost revenues and create additional revenues (e.g., new business models) for service providers. In some scenarios, smart contracts could trigger actions (e.g., reimbursements) based on the data collected (e.g., from physical sensors) or ease identity verification. In the context of fraud prevention, the blockchain could act as a global shared ledger record, including for example, a person's previous history (e.g., previous claims, traffic violations).

## VI. CONCLUSIONS

The transition to a data and value-driven world is fostered by the pace of the technological disruptions of an Internet-enabled global world, the challenges of future mobility, and an increasing business competition. In this ever more complex ecosystem, the use of blockchain can provide to the automotive industry a platform able to distribute trusted and cyber-resilient information that defy current non-

**IEEE** *Access*

collaborative organizational structures. It must be noted that despite the hype reported by numerous organizations, it is necessary to perform an objective evaluation about how and whether to invest or not in blockchain from a business management and cybersecurity standpoint.

This article covered a broad suite of issues that arise from the advent of a disruptive technology like blockchain. In addition, we present a holistic approach to a blockchain-based advanced automotive industry with a review of the main scenarios and the optimization strategies for designing and deploying these applications. Furthermore, some recommendations were mentioned to guide future researchers and managers on some of the open issues that will have to be confronted before deploying the next generation of secure blockchain applications.

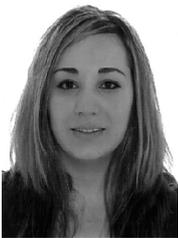

PAULA FRAGA-LAMAS (M'17) received the M.Sc. degree in Computer Science in 2008 from University of A Coruña (UDC) and the M.Sc. and Ph.D. degrees in the joint program Mobile Network Information and Communication Tech- nologies from five Spanish universities: University of the Basque Country, University of Cantabria, University of Zaragoza, University of Oviedo and University of A Coruña, in 2011 and 2017, re- spectively. Since 2009, she has been working with the Group of Electronic Technology and Communications (GTEC) in the Department of Computer Engineering (UDC). She holds an MBA and postgraduate studies in business innovation management (JMC of European Industrial Economy), sustainability (CSR) and social innovation (INDITEX- UDC Chair). She is co-author of more than fifty peer-reviewed indexed journals, international conferences and book chapters. Her current research interests include wireless communications in mission-critical scenarios, Industry 4.0, Internet of Things (IoT), Augmented Reality (AR), blockchain, RFID and Cyber-Physical systems (CPS). She has also been participating in more than twenty research projects funded by the regional and national government as well as R&D contracts with private companies.

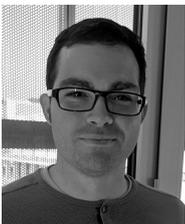

TIAGO M. FERNÁNDEZ-CARAMÉS (S'08- M'12-SM'15) received his MSc degree and PhD degrees in Computer Science in 2005 and 2011 from University of A Coruña, Spain. Since 2005 he has worked as a researcher and professor for the Department of Computer Engineering of the University of A Coruña inside the Group of Elec- tronic Technology and Communications (GTEC). His current research interests include IIoT/IoT systems, RFID, wireless sensor networks, Industry 4.0, blockchain and augmented reality.

...